\documentclass{aa}  
\usepackage{graphicx}
\usepackage{amsmath}
\usepackage{amssymb}
\usepackage{natbib}

\usepackage{txfonts}
\newcommand{\vl}{{\varv}}
\newcommand{\pr}[1]{\ensuremath{{#1}_{\mathrm{p}}}}
\newcommand{\io}[1]{\ensuremath{{#1}_{\mathrm{i}}}}
\newcommand{\prio}[1]{\ensuremath{{#1}_{\mathrm{pi}}}}
\newcommand{\kpi}{\ensuremath{\prio{k}}}
\newcommand{\xpi}{\ensuremath{\prio{x}}}
\newcommand{\xpio}{x_{\rm pi,0}}
\newcommand{\xpii}{x_{\rm pi,1}}
\newcommand{\Rpi}{\ensuremath{\prio{R}}}
\newcommand{\vp}{\ensuremath{\pr{\vl}}}
\newcommand{\vi}{\ensuremath{\io{\vl}}}
\newcommand{\Ap}{\ensuremath{\pr{A}}}
\newcommand{\Ai}{\ensuremath{\io{A}}}
\newcommand{\qi}{\ensuremath{\io{q}}}
\newcommand{\qp}{\ensuremath{\pr{q}}}
\newcommand{\vth}{\ensuremath{{\vl}_\mathrm{th}}}
\newcommand{\geff}{\ensuremath{g_\mathrm{eff}}}
\newcommand{\pderiv}[2]{\ensuremath{\frac{\partial #1}{\partial #2}}}
\newcommand{\deriv}[2]{\ensuremath{\frac{\mathrm{d} #1}{\mathrm{d} #2}}}
\newcommand{\derivl}[2]{\ensuremath{{\mathrm{d} #1}/{\mathrm{d} #2}}}
\newcommand{\sigeref}{\ensuremath{\sigma^\mathrm{ref}_\mathrm{e}}}
\newcommand{\Rstar}{\ensuremath{R_\ast}}
\newcommand{\vinfty}{\ensuremath{\vl_\infty}}
\newcommand{\vinftyp}{\ensuremath{\vl_{\infty,\rm p}}}
\newcommand{\vinftyi}{\ensuremath{\vl_{\vinfty,\rm i}}}
\newcommand{\vinftypol}{\ensuremath{\vl^{3}_\infty}}
\newcommand{\gb}{\ensuremath{\Gamma_{\rm B}}}
\newcommand{\gl}{\ensuremath{\Gamma_{\rm L}}}
\newcommand{\rd}{\ensuremath{r_{\rm d}}}
\begin{document}

\title{A hydrodynamic scheme for two-component winds from hot stars}

\author{V. Votruba\inst{1,}\inst{3},
	A. Feldmeier\inst{2},
	J. Kub\'{a}t\inst{1},
	D. R\"{a}tzel\inst{2}}

\authorrunning{V. Votruba et al.}

\offprints{V. Votruba}

\institute{Astronomick\'{y} \'{u}stav, Akademie v\v{e}d \v{C}esk\'{e}
	republiky, CZ-251 65 Ond\v{r}ejov, Czech Republic \\
	\email{votruba@sunstel.asu.cas.cz}
	\and
	Astrophysik, Institut f\"ur Physik, Universit\"at Potsdam, Am
	Neuen Palais 10, D-14469 Potsdam, Germany\\
	\email{afeld@astro.physik.uni-potsdam.de} 
	\and
	\'Ustav teoretick\'e fyziky a astrofyziky,
	P\v{r}irodov\v{e}ck\'{a} fakulta, Masarykova univerzita,
	Kotl\'{a}\v{r}sk\'{a} 2, CZ-611 37 Brno, Czech Republic}

\date{Received 11 September 2006}

\abstract
{}
{We have developed a time-dependent two-component
hydrodynamics code to simulate radiatively-driven stellar winds from hot stars.}
{We use a time-explicit van Leer scheme to solve the hydrodynamic
equations of a two-component stellar wind. Dynamical friction due to
Coulomb collisions between the passive bulk plasma and the
line-scattering ions is treated by a time-implicit, semi-analytic
method using a polynomial fit to the Chandrasekhar function. This
gives stable results despite the stiffness of the problem.}
{This method was applied to model stars with winds that are both poorly and well-coupled. While for the former case we reproduce the mCAK solution,
for the latter case our solution leads to wind decoupling.}
{}
\maketitle


\section{Introduction}

Stellar winds from hot stars are often described using the so-called
CAK theory \citep*{C1975}, including later improvements for a finite
stellar disk \citep{{Pa1986},{rdsw3}} and the ionisation
stratification of the wind \citep{rdsw2}. The radiative line force is
here parameterised by three constants only, $k,\alpha$, and $\delta$.

Although qualitative agreement between theory and observations was
achieved, there were still discrepancies remaining, namely the
terminal velocity {\vinfty} being too low and the mass-loss rate
$\dot{M}$ being too high \citep{rdsw2,rdsw3}. {Besides this,
calculations did not reproduce the observed ionisation ratios
\citep{PP1990}. Further development of the theory was focused mainly
on the effect of multiline scattering, magnetic field and rapidly
rotating stars. As shown by \cite{rdsw4}, if one includes the
effect of multiple scatterings in overlapping lines, the terminal
velocity is higher than the CAK value, while the mass-loss rate
remains similar to the latter.  Furthermore, the inclusion of a
magnetic field and the assumption of high rotational
velocity lead to a higher terminal velocity and a higher mass-loss
rate \citep{rdsw5}.

The above problem with the ionisation structure can be resolved
by including X-rays. X-rays are a common feature among the O
stars and can change the ionisation structure of the
wind significantly. As was first discussed by \citep{CO1979} and later by
\citep{PP1990}, the X-rays lead to photoinisation via Auger processes
and are necessary to explain the presence of superionised ions {O VI}
in observations.

\citet{LS70} suggested a new hydrodynamic instability of the radiative
driving force. This so-called line-driven instability can possibly
explain observed variabilities of stellar winds, such as variable
X-ray emission, discrete absorption components, and the appearance of
broad absorption troughs in P Cygni line profiles \citep{Pr1986}. The
contradictory results from the simplified linear stability analysis of
\cite{MHR79} and \cite{C1980}, on one hand, and \cite{A80}, on the
other, were unified by \cite{O1984,O1985} in the so-called
bridging law. The first time-dependent, numerical simulations of the
instability were undertaken by \cite{O1988}, who found that the wind
develops a train of strong reverse shocks. Their model was improved by
\cite{Fe1995}, including the energy equation without the approximation
of gas isothermality to calculate the temperature structure
of the stellar wind. Adopting a simple formulation for the line force,
\cite{O2005} extended hydrodynamical instability simulations from one
to two dimensions (2D), to find the lateral coherence length of
instability-generated shells of dense gas in the wind.

The approximation of the one-component flow assumed in the CAK model is
acceptable for sufficiently dense winds from O stars and hot B stars
\citep*{CAK76}. The radiative line force acts only on ions (termed
``metals'' in the following) that scatter photons in numerous spectral
lines. These ions share their momentum through Coulomb collisions with
the passive part of the plasma (protons). The dynamical effect of
Coulomb collisions on the plasma is described well by dynamical
friction, which was introduced by \cite{Chandrasekhar} for the
gravitational force and later applied to electromagnetic forces by
\cite{Spitzer}.

For thin winds from B\,near-main-sequence stars, however, decoupling
and a well-known plasma runaway effect can occur \citep{S1992}. At low
wind densities, Coulomb interactions are weak, and the momentum
transfer from metals to protons becomes inefficient. As a result, the
two components decouple at a given radius. From this decoupling
radius on, metals are strongly accelerated by the radiative force,
whereas the passive plasma is decelerated by gravity.

It came as a surprise that \cite{KK00} obtained a different result
from stationary hydrodynamic calculations. They found a wind solution
where the two components remain coupled, but undergo a sudden jump to
a slow-acceleration branch at a definite radius. This is analogous to
what is found for shallow (and overloaded) solutions \citep{Fe2000}.
The issue has not been settled yet ever since \cite{O2002} showed from linear
stability analysis that the flow should be disrupted by ion separation
before reaching a slow-acceleration solution. \cite{KK02} also
found that in a low-density wind, frictional heating may be important
for the energy balance of the whole wind and may increase the
temperature of the gas significantly.  

An effect similar to ion decoupling was considered by
\cite{P1999}, namely shock decoupling.  In a low-density wind that
passes through a shock, the postshock gas remains at high temperatures
and in a highly ionised state because the gas is too rarefied to cool
radiatively \citep{KR85}. In this case the ions responsible for line
driving are completely stripped and are not accelerated further by the
radiation field.  

Another interesting idea in this context is the generation of
pulsating shells \citep{P1999}. If the wind decouples at a radius where
the flow speed is still lower than the escape speed, the passive
plasma falls back to the star, and the interaction with the outflow
leads to pulsating shells. However, this result is based only on a
one-component model, with an artificial turn-off of the radiative
acceleration.

To analyse the possible occurrence of ion decoupling and pulsating
shells in thin stellar winds, it is necessary to develop a suitable
time-dependent, two-component hydrodynamics code. In the present paper
we describe our numerical method to simulate these outflows, and give
basic results on thick and thin winds.

\section{Two-component winds}

We restrict ourselves to a 1D spherically symmetric, isothermal,
quasineutral, two-component outflow consisting of metals that scatter
stellar photons in numerous spectral lines and a passive plasma. The
forces acting are gravity, dynamical friction, gas pressure gradients,
and, for only the line-scattering ions, the radiative line force. The
continuity equations are
\begin{subequations}
\label{kont}
\begin{eqnarray} 
\pderiv{\pr\rho}{t}+
\frac{1}{r^2}\pderiv{(r^2 \pr\rho \vp)}{r}&=&0\,, \\
\pderiv{\io\rho}{t}+
\frac{1}{r^2}\pderiv{(r^2 \io\rho \vi)}{r}&=&0\,,
\end{eqnarray}
\end{subequations}
and the Euler equations are
\begin{subequations}
\label{momentum}
\begin{eqnarray}
\pderiv{\vp}{t}+\vp\pderiv{\vp}{r}+\frac{1}{\pr\rho}\pderiv{\pr{p}}{r}
	&=&\frac{\Rpi}{\pr\rho}-\geff \,, \\		
\pderiv{\vi}{t}+\vi\pderiv{\vi}{r}+\frac{1}{\rho_i}\pderiv{\io{p}}{r}
	&=& g_{\rm rad}^{i}-\geff-\frac{\Rpi}{\io\rho} \,.
\end{eqnarray}
\end{subequations}
Here, {\io\rho}, {\vi}, and $\io{p}$ are the density, velocity, and
pressure of the metals, and {\pr\rho},{\vp}, and $\pr{p}$ are the
density, velocity, and pressure of the passive plasma. The effective
gravitational acceleration is $\geff = -GM_*(1-\Gamma_e)/r^2$, with
gravitational constant $G$ and Eddington factor $\Gamma_e$. The
frictional force {\Rpi} between metals and passive plasma is described
below (see Eq.\,\ref{trsila}).  The above system is closed by the
equations of state,
\begin{subequations}
\begin{align}
\pr{p}&=\sqrt{kT/\pr{m}}\pr\rho, \\
\io{p}&=\sqrt{kT/\io{m}}\io\rho.
\end{align}
\end{subequations}
The assumption of wind isothermality becomes questionable in the presence
of strong frictional heating, and we aim to include the energy
equation in future work.

\subsection{Radiative acceleration}

We write the radiative acceleration in the form \citep[Eq.3]{KK00}
\begin{equation}\label{zazzry}
g_\mathrm{rad}^{i}(r)=\frac{(\eta\sigma_e)^{1-\alpha}}{4\pi\vth^\alpha}
\frac{L_*}{r^2}k\left(\frac{1}{\io\rho}\pderiv{\vi}{r}\right)^{\alpha}
f_\mathrm{ion}f_\mathrm{fin},
\end{equation}
with CAK force multipliers $k,\alpha,\delta$. Here, $L_*$ is the
luminosity of the star, {\vth} the thermal velocity of ions, and
$\sigma\approx 0.33$ cm$^2$ g$^{-1}$ the Thomson opacity due to
scattering on electrons.

The radiative acceleration due to line scattering only acts on metal ions, but the CAK force multipliers $k,\alpha,\delta$ were calculated
for a one-component plasma. We account for this by a scaling factor
$\eta$ to the radiative force. The value $\eta=0.0127$ calculated by
\citet{KK00} for solar metallicities is adopted here.

The finite disk correction factor $f_{\rm fin}$ is given by
\citep{C1975}
\begin{equation}
f_{\rm fin}(r)=\frac{{(1+\sigma)}^{\alpha+1}-(1+\sigma\mu_*^2)}
{\sigma(\sigma+1)^{\alpha}{(1-\mu_*^2)}{(\alpha+1)}} \,,
\end{equation}
where $\mu_*=\sqrt{1-(\Rstar/r)^2}$ (with stellar radius $\Rstar$),
and $\sigma$ is given by \citep{C1974}:
\begin{equation}
\sigma=\frac{r}{\vi}\pderiv{\vi}{r}-1\,.
\end{equation}
Finally, $f_{\rm ion}$ is a correction for the ionisation state of the
stellar wind \citep{rdsw2},
\begin{equation}
f_{\rm ion}(r)={\left(\frac{10^{-11}[{\rm
cm}^3]\,n_\mathrm{e}}{W(r)}\right)}^{\delta} \,,
\end{equation}
where $W(r)=\frac{1}{2}(1-\sqrt{1-(\Rstar/r)^2})$ is the geometrical
dilution factor. Due to the assumption of quasineutrality, the number
density of electrons $n_e$ roughly matches the number density of the
passive plasma. Thus, we use $n_e=n_p$ in $f_{\rm ion}$. This
correction factor does not have a significant influence on the wind
dynamics \citep{rdsw2}.

\subsection{Friction terms} 

The passive plasma and absorbing ions interact via Coulomb collisions,
which are described by a frictional force {\Rpi} per volume
\citep{S1992},
\begin{equation}
\label{trsila}
\Rpi=\pr{n} \io{n} \kpi G(\xpi)\,,
\end{equation}
where $\pr{n}$ and $\io{n}$ are the number densities of the passive
plasma and absorbing ions, respectively, and the frictional
coefficient {\kpi} is given by
\begin{equation}
\kpi=\frac{4\pi \ln{\Lambda} \pr{Z}^2 \io{Z}^2 e^4}
{k_\mathrm{B}T}
\frac{\vi-\vp}{|\vi-\vp|}.
\end{equation}
Here, $\io{Z}e$ and $\pr{Z}e$ are the ion and passive plasma charges,
respectively. The Coulomb logarithm $\ln{\Lambda}$ is defined as
\begin{equation}
\ln{\Lambda}=\ln{\left[\frac{24\pi}{\sqrt{n}}
{\left(\frac{k_{B}T}{4\pi e^2}\right)^{1.5}}\right]},
\end{equation}
with Boltzmann constant $k_{B}$, total number density $n$, and wind
temperature $T$. The Chandrasekhar function $G(x_{\rm pi})$
\citep{Chandrasekhar,Spitzer} in Eq.\,(\ref{trsila}) is given in terms
of the error function $\Phi(\xpi)$ by \citep{Spitzer}
\begin{equation}
\label{chpresna}
G(\xpi) = \frac{\Phi(\xpi)}{2 \xpi^2}
-\frac{\exp{(-\xpi^2)}}{\xpi\sqrt{\pi}}.
\end{equation}
This function depends on {\xpi}, the ion separation drift speed
relative to the passive plasma, scaled to the mass-weighted thermal
speed \citep{S1992}
\begin{equation}
\xpi=\frac{|\vi-\vp|}{\vth\sqrt{1+\io{A}/\pr{A}}},
\end{equation}
where $\io{A}$ and $\pr{A}$ represent the mean atomic mass of ions and
passive plasma in atomic units.

\section{Method of solution}

To solve the four hydrodynamic equations we use a
hydrodynamics code developed by \citet{Fe1995} as a core. It employs a standard
Euler scheme and is suited for 1D, one-component
outflows. We extended this code to a two-component version. Equations
(\ref{kont}) and (\ref{momentum}) are discretised using an
operator-splitting, time-explicit, finite difference method on a
staggered mesh \citep[see][p.~131]{leveque}. We calculate advection
fluxes using van Leer's monotonic interpolation
\citep[see][]{vanLeer}.

\cite{KK00} discussed the circumstances under which the radiative line
acceleration is balanced by dynamical friction and not by inertia. In
this case, a decrease in dynamical friction leads
(counter-intuitively) to a decrease in the radiative acceleration of
the gas. We accelerate the gas by the \emph{sum} of dynamical friction
and radiative force, i.e.~avoid operator-splitting of these two force
terms, in order to achieve a stable numerical scheme. Similarly, for a
barometric density stratification, one has to apply the sum of the
thermal pressure force and gravity at once to avoid an unstable
scheme.

As the time step, we use the minimum of the separate Courant time steps for
the two flow components. The wind is characterised by two parameters,
the mass-loss rate and terminal speed. The latter scales with the
escape speed, and the former is given roughly by the CAK relation
\cite[Eq.\,46]{C1975}
\begin{equation}
\dot{M}_{\rm CAK}=
\frac{4\pi G M}{\sigeref \vth}
\alpha(1-\alpha)^{(1-\alpha)/\alpha}
(k\Gamma_e)^{1/\alpha}(1-\Gamma_e)^{-(1-\alpha)/\alpha}.
\label{CAKMdot}
\end{equation}
To calculate the radiative force we use tabulated values of the CAK
parameters from \cite{rdsw2}.

As initial conditions for the case of a well-coupled wind we use, for
the velocity of the both components in the subsonic part, $\vl(r)=0.1
{a}\exp{\left(H r\right)}$, where $H$ is the scale height and $a$ the
isothermal sound speed. In the supersonic part, we use $\vl(r)=a+20\,
a\, r\,$. Initial values for the density of both components are
calculated from Eqs.\,\eqref{kont} and \eqref{CAKMdot}.

For the case of a low-density wind (where poor coupling between the
components may be expected), we artificially increase the frictional
coupling by increasing the average ion charge $\qi$ and run the
simulation (with the above initial conditions) until a converged CAK
one-component solution is achieved. This solution then serves as
initial conditions for both components in a subsequent simulation with a
realistic average ion charge.

Our boundary conditions are set according to the theory of
characteristics \citep[pp.~303--307]{anderson}. For absorbing ions and
passive plasma, we keep the densities $\pr\rho$ and $\io\rho$ at the
inner boundary fixed to their initial values and set, at each time
step, the momentum densities $\io\rho\vi$, $\pr\rho\vp$ at the inner
boundary to their value at the first interior mesh point (zero-order
extrapolation). At the outer boundary, we extrapolate the mass and
momentum densities of both components from the last interior mesh
point.

\subsection{Discretisation of friction}

The central issue of the present paper is the inclusion of the
frictional term \eqref{trsila}. In the following we develop a simple,
computationally effective, and accurate method for determining
of the velocity difference {\xpi} at every time step, based on the
frictional term.

\begin{figure}[ht]
   \begin{center}
    {\includegraphics{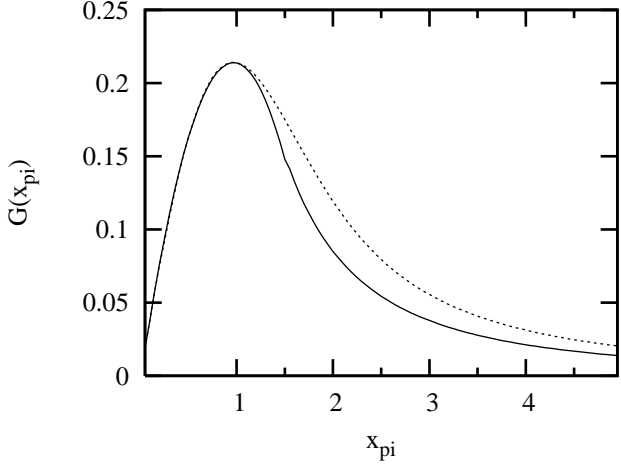}}
   \end{center}
\caption{The Chandrasekhar function: exact function $G(\xpi)$
according to \eqref{chpresna} (dashed line), and approximation
$G_A(\xpi)$ (solid line).}
\label{Chandra}
\end{figure}

The equations for the momentum change of the passive plasma and the
absorbing ions due to the frictional term alone are, from
\eqref{momentum}
\begin{align}
\label{Chandrasplit1}
\frac{\partial {\vl_{r,i}}}{\partial t} &=
-\frac{\mathcal{C}G(\xpi)}{\rho_p}, \\
\label{Chandrasplit2}
\frac{\partial {\vl_{r,p}}}{\partial t}&=
\frac{\mathcal{C}G(\xpi)}{\rho_i},
\end{align}
where the constant $C$ is given by
\begin{equation}
\mathcal{C}=-\frac{\kpi}{\Ai \Ap m_p^2}.
\end{equation}

Transforming the above differential equations into time-explicit
difference equations results in a system with high stiffness, since we
attempt to evolve a stellar wind on a relatively slow flow time scale
with respect to which the faster, frictional processes maintain
equilibrium almost instantaneously. As a consequence, the numerical
solution fails due to large oscillations at the wind base. Therefore
we set up the following semi-implicit scheme for the frictional
terms. Subtracting Eq.\,\eqref{Chandrasplit2} from
Eq.\,\eqref{Chandrasplit1}, we obtain a differential equation for the
drift velocity,
\begin{equation}
\label{driftrych}
\deriv{\xpi}{t}=-\frac{(\rho_i+\rho_p)
\mathcal{C}}{\vth\sqrt{1+\Ai/\Ap}}G(\xpi).
\end{equation}
In Eqs.\,(\ref{Chandrasplit1}) and (\ref{Chandrasplit2}) we now replace the
partial time derivatives by a total derivative following the idea of
operator-splitting, i.e.\, ,friction is now treated without reference to any
``forthcoming'' hydrodynamic processes. The $d/dt$ is not to be
understood as a Lagrange derivative. Upon formal integration,
\begin{equation}
\int_{\xpio}^{\xpii}\frac{{\rm d}\xpi}{G(\xpi)}=
-\int_{t_0}^{t_0+\Delta t}
\mathcal{C}\frac{(\rho_i+\rho_p)}{\vth\sqrt{1+A}}
{\rm d}t,
\label{Chandraint}
\end{equation}
where $A$ denotes the ratio $\Ai/\Ap$, and $\xpio$, resp.~$\xpii$, are
the drift velocities at times $t_0$, resp.~$t_0+\Delta t$. We want to
obtain an \emph{analytical} expression for the left hand side of
Eq.\,\eqref{Chandraint}. To this end we use a three-interval fit to
the Chandrasekhar function. For low and high drift velocities we
use the same approximations as \citet{O2002},
\begin{equation}
G_1(\xpi) \approx \frac{2\xpi}{3\sqrt{\pi}} \quad {\mathrm{for}} \,
\xpi\leq x_1
\end{equation}
and
\begin{equation}
G_3(\xpi) \approx \frac{K}{2\xpi^2} \quad {\mathrm{for}} \,
\xpi\geq x_2.
\label{Scalling}
\end{equation}
For values between $x_1$ and $x_2$, we approximate the Chandrasekhar
function using a quadratic function,
\begin{equation}
G_2(\xpi) \approx a_2\xpi^2+b_2\xpi+c_2,
\label{ch2}
\end{equation}
where the parameters $a_2,b_2,c_2$ are evaluated from the conditions
$G_1(x_1) = G_2(x_1)$, $G_2(x_{\rm max}) = 0.214$, and
$\derivl{G_1(x_1)}{x} = \derivl{G_2(x_1)}{x}$.  Continuity of the
derivative at $x_2$ is not required, because we have only a second-order 
polynomial approximation. The points $x_1,x_2$, and the 
scaling factor $K$ were chosen to achieve the best fit to the
Chandrasekhar function \eqref{chpresna}. We find, for $x_1=0.1$ and
$x_2=1.5$, the values $a_2 = -0.2341$, $b_2 = 0.4532$, $c_2 =-0.0053$
and $K=0.74$. This approximation to the Chandrasekhar function is
termed $G_A$ in the following and is shown in Fig\,\ref{Chandra}.

For a well-coupled wind the difference between $G$ and $G_A$ is of
minor importance, and both functions lead to essentially the same wind
solution. To show this, we use two different approximations $G_A$, one
being a global overestimate, the other a global underestimate of
$G$. In both cases, the code converged to almost the same steady CAK
solution.

Using the approximation $G_A$, we obtain the velocity difference {\xpi}
as follows. At the dynamically most important part of the
Chandrasekhar function, i.e.~its maximum (covered by $G_2$),
integration of \eqref{Chandraint} leads to
\begin{equation}
\ln \left.\frac{2a_2\xpi+b_2-\Xi}{2a_2\xpi+b_2+\Xi}\right|^{\xpii}_{\xpio}=
-\mathcal{C}\Delta t \,\Xi,
\label{diskretchandterm}
\end{equation}
where $\Xi=\sqrt{b_2^2-4a_2c_2}$. With the help of the substitutions
\begin{align}
q_{-} &= 2a_2\xpio+b_2-\Xi, \\
q_{+} &= 2a_2\xpio+b_2+\Xi, 
\end{align}
we can simplify the expression \eqref{diskretchandterm} and as
final form of the velocity difference obtain %
\begin{equation}
\label{anal2}
\xpii=\frac{1}{2a_2}
\left[\frac{1+\frac{q_{-}}{q_{+}}\exp{\left(-\mathcal{C}\Delta t
\Xi\right)}}{1-\frac{q_{-}}{q_{+}}\exp{\left(-\mathcal{C}\Delta t
\Xi\right)}}\Xi- b_2 \right].
\end{equation}
For low drift velocities and using the approximation $G_1(\xpi)$,
integration of \eqref{Chandraint} gives
\begin{equation}
\label{anal1}
\xpii=\xpio\exp{\left(-\frac{2}{3\sqrt{\pi}}\mathcal{C}\Delta
t\right)},
\end{equation}
and finally, for high drift velocities and using the {approximation}
$G_3(\xpi)$,
\begin{equation}
\label{anal3}
\xpii=\sqrt[3]{\xpio^3-\frac{3}{2}K\mathcal{C}\Delta t}.
\end{equation}
The expressions (\ref{anal2},\ref{anal1},\ref{anal3}) are used in our
code to calculate the change in drift velocity {\xpi} due to friction
during a hydrodynamic time step. This results in a highly improved
stability behaviour compared to direct, time-explicit differencing of
the friction terms and, for the first time, allows time-dependent
simulations of multi-component winds.

Finally, we note that a commonly used approximation to the
Chandrasekhar function,
\citep{Karlicky}
\begin{equation}
\label{wholeaprox}
G_A(\xpi)= \frac{K \xpi}{(\xpi^2+\vth^2)^{3/2}},
\end{equation}
which covers the whole interval of drift speeds, results in integrals
we were not able to carry out analytically.

\subsection{Time requirements}

Since the two-component code is essentially a duplication of the
one-component code, and since the frictional force terms are solved by
analytic equations, i.e.\,at very small computational cost, the cpu
time for a simulation is comparable to the one-component case,
typically a few hours (for $O(10^3)$ mesh points) on a dedicated
workstation. However, we chose a very small Courant number of 0.05
(instead of typically 0.5) to bring the inviscid Courant time step
somewhat closer to the frictional time step, in order to avoid changes too
large in friction terms during subsequent time steps, which
could trigger instability.


\section{Results of calculations}

\begin{figure*}[t!]
\scalebox{1.0}{\includegraphics{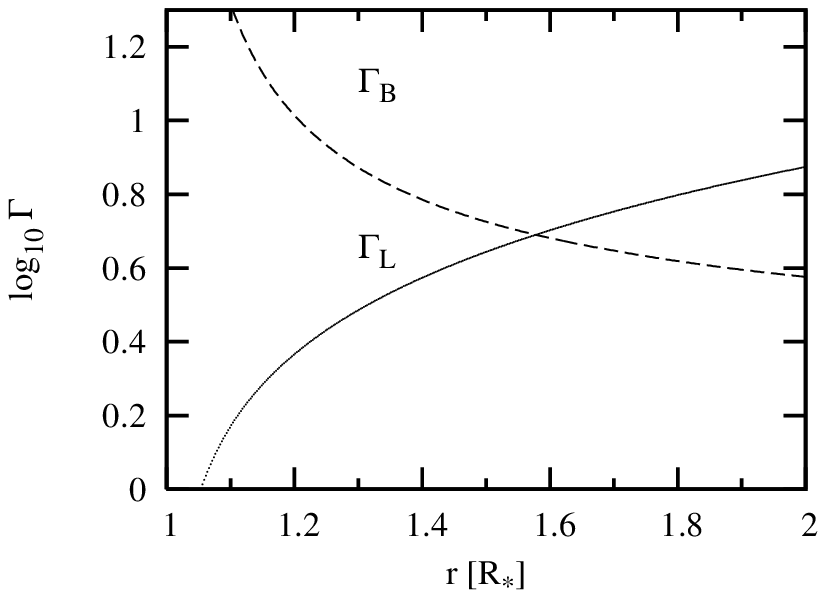}}
\scalebox{1.0}{\includegraphics{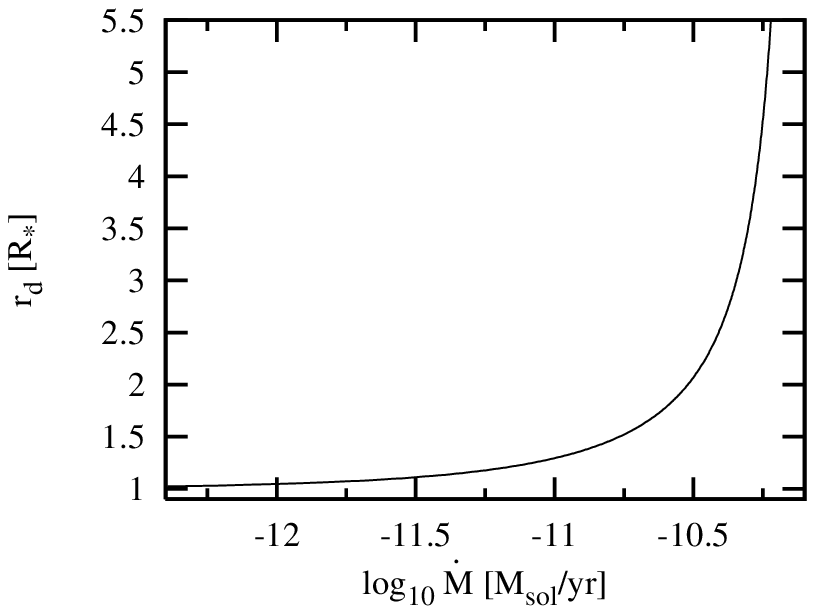}}
\caption{{\em Left panel:} Plot of $\gb$ and $\gl$ as function of
radius for the model B5. The curves cross at $r \sim 1.6\,\Rstar$.  {\em
Right panel:} Dependence of the decoupling radius $\rd$ on mass-loss
rate $\dot{M}$ for the model B5. If the wind is denser, the decoupling
radius is located farther away from the star.}
\label{poorlycoupled}
\end{figure*}

In the present paper, we considered essentially the same stars as
\citet{KK00}. More specifically, we apply our method to the wind where
a well-coupled solution is predicted (we refer to this case as B0)
and to the wind with possible decoupling (referred to as B5).
The corresponding stellar parameters are given in Table
\ref{Modely}. Details about these parameters are discussed by
\cite{KK00}.

To determine whether our model winds are well or poorly coupled, we
considered the parameter $\gl$, which is the ratio of the radiative
force per mass acting on ions, $g_\mathrm{rad}^{i}$, to the
gravitational acceleration $g$ \citet{S1992},
\begin{equation}
\gl=\frac{\io\rho}{\io\rho+\pr\rho}\frac{g_\mathrm{rad}^{i}(r)}{g}=
\frac{g_{\mathrm{rad}}^{\mathrm CAK}}{g},
\end{equation}
and the parameter $\gb$, which is the ratio of the CAK radiative force
$g_{\mathrm rad}^{\mathrm CAK}$ to gravity, for the case that
dynamical friction reaches a maximum,
 
\begin{equation}
\gb=\frac{\left. g_{\mathrm{rad}}^{\mathrm CAK}\right|_{\xpi=0.968}}{g}.
\end{equation}

Decoupling should occur, if
\begin{equation}
\gb < \gl.
\end{equation} 
>From Fig.\ref{poorlycoupled}, decoupling is expected in model B5 for
$r\gtrsim 1.6\Rstar$. Analytically, $\rd$ is determined by
\citet{S1992},
\begin{equation}
\rd=R_*{\left(1-{\left(\frac{\dot{M}_{\mathrm CAK} \eta \, \kpi
G_{\mathrm max}}{4\pi\,\beta\,{R_*}\vinftypol} \right)}^{{1}/{(3
\beta-1)}}\right)}^{-1},
\label{decouplerd}
\end{equation}
where $\beta$ is the beta-law parameter \cite[see, e.g.,][]{P1999}
and $\eta$ has been introduced in \eqref{zazzry}.

\subsection{Well coupled stellar wind}
\begin{figure*}[t!]
\scalebox{1.0}{\includegraphics{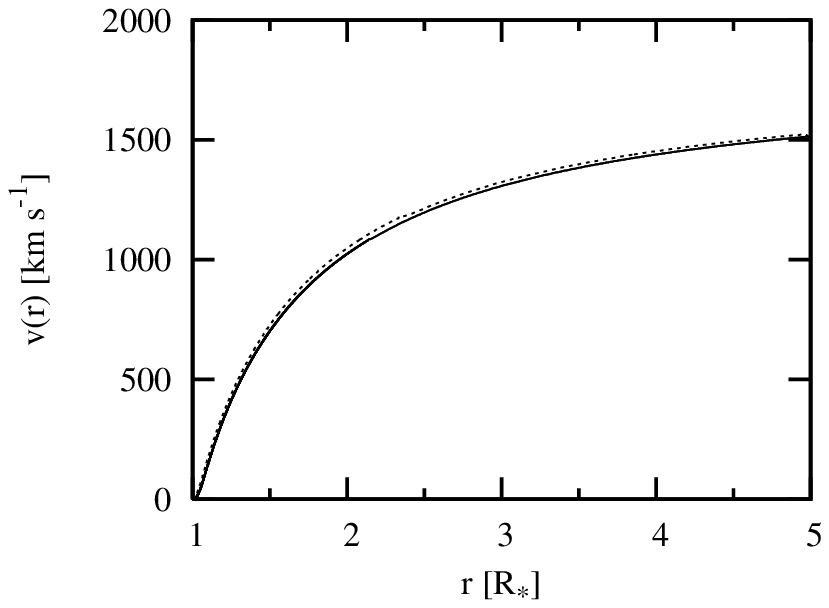}}
\scalebox{1.0}{\includegraphics{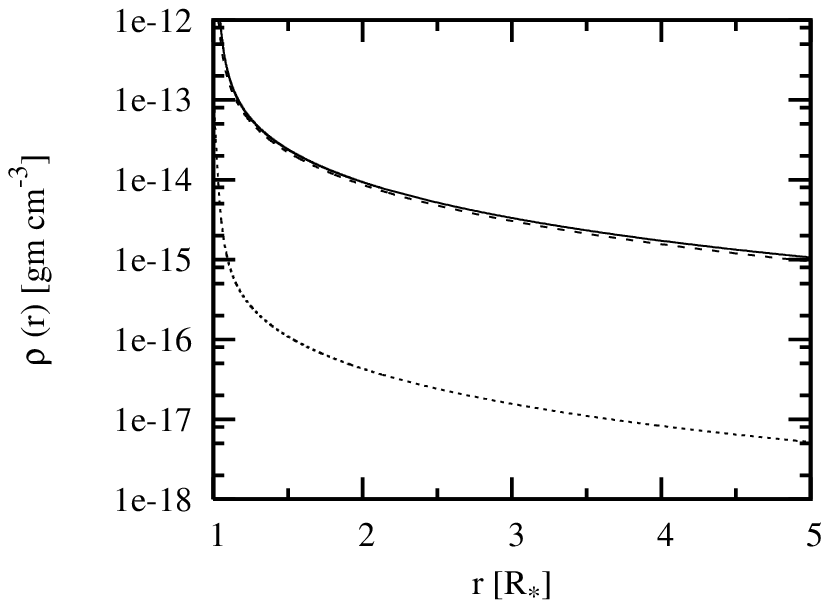}}\\
\caption{{\it Left {\bf \em panel:} } Final wind-velocity laws for
model B0 after 55 flow time units $R_*/\vinfty$.  Absorbing ions are
marked by (---), passive plasma by ( - - - ), and the result of the
one-component code by ($\cdots$). The curves for ions and passive
plasma are indistinguishable. {\it Right {\bf \em panel:}} Density for
the same model B0.}
\label{B0Solution}
\end{figure*}

For this case we used the B0 model star from \cite{KK00}. Our
hydrodynamic simulation evolved to a steady mCAK solution with the
correct terminal speed $\vinfty$. Figure \ref{B0Solution} shows the
velocity law and the density stratification of the wind. For
comparison with \cite{KK00} and the CAK predictions, we summarise the
results of our calculation in Table~\ref{Vysledky}. As was expected,
for the case of a well-coupled wind, the velocities of both absorbing
ions $\vinftyi$ and passive plasma $\vinftyp$ are roughly the same.
\begin{table}[b]
\caption{Parameters of wind models}
\label{Modely}
 \begin{tabular}{c c c c c c c c}
 \hline
Star & $ M $ &  $T_{\rm eff}$ & \Rstar & $\alpha$ & k & $\delta$ & \qi/\qp\\
 & [M$_{\odot}$] & [K] & [R$_{\odot}$] & & & &\\
\hline\hline
  B0 & 90.0 & 28500 & 37.0 &  0.590 & 0.170 & 0.09 & 3.0 \\
  B5 & 4.36 & 15500 & 3.01 &  0.511 & 0.235 & 0.12 & 2.5 \\
\hline
\end{tabular}
\label{modeltable}
\end{table}

\begin{table}[h]
\caption{Calculated values of $\vinftyp$ and $\vinftyi$ in comparison
with $\vinfty^\mathrm{CAK}$ from the one-component model, and derived
values of the mass-loss rate for absorbing ions, $\dot{M}_{\rm
i}$, and for the passive plasma, $\dot{M}_{\rm p}$.}
\label{Vysledky}
\begin{tabular}{c c c c c c}
\hline
Star & \vinftyp  & \vinftyi & $\vinfty^{\rm CAK}$ & $\dot{M}_{\rm i}$ & $\dot{M}_{\rm p} $ \\
& [km.s$-1$] &[km.s$-1$] & [km.s$-1$] & [$\dot{M}_{\odot}$ p. y.] & [$\dot{M}_{\odot}$ p. y.] \\
\hline\hline
B0 & 1600 & 1600 & 1600 & 6.6$\, 10^{-8}$ & 4.4$\, 10^{-6}$  \\
B5 &  200 & 36000 & 800 & 2.3$\, 10^{-14}$ & 1.5$\, 10^{-12}$\\
\hline
\end{tabular}
\note{The value of the passive plasma mass-loss rate roughly corresponds to
the CAK value.}
\end{table}

The agreement between our results and those obtained with the
one-component version of the code is very good, see
Fig.~\ref{B0Solution}.  There is also good agreement with the
stationary calculation by \cite{KK00}.

To test the sensitivity of the results to our approximation of the
Chandrasekhar function, we used different sets of parameters
$a_2,b_2,c_2$ and of points $x_1,x_2$ without the condition of a best
fit. For $a_2=-0.176$, $b_2=0.404$, $c_2=-0.001$, $x_1=0.15$, and
$x_2=1.85$, fit $G_A$ lies everywhere above $G$, with maximum
deviation $\approx 30\%$, whereas for $a_2=-0.224$, $b_2=0.438$,
$c_2=-0.0033$, $x_1=0.10$, and $x_2=1.10$, the function $G_A$ lies
everywhere below $G$, with maximum deviation $\approx 50\%$. It turns
out that, for the case of a well-coupled wind, the detailed form of the
approximation of the Chandrasekhar function is not very important,
since the steady-state wind solutions are always fairly similar. On
the other hand, for a poorly coupled wind, a correct approximation is
mandatory, since the frictional force determines the point where the
wind starts to decouple.

\subsection{Low-density wind with decoupling}

As a second model, we considered the B5 star from \cite{KK00}. Their
two-component model shows a solution with lower acceleration
compared to the normal CAK solution, but did not lead to decoupling.
Results of our calculations are shown in Fig.~\ref{B5}.
Compared to \cite{KK00}, we changed the average ion charge to a
slightly higher value $\qi=2.5$ instead of $\qi=2.0$, as before.  The
aim of this was to increase dynamical friction to prevent
the appearance of pulsating shells. Namely, this higher value of $\qi$
 shifts the decoupling radius to a location where the escape velocity
is lower than the local speed of the passive plasma and,
consequently, matter is no longer gravitationally bound to the star.
The more subtle case when the matter is still gravitationally bound to
the star will be considered in a forthcoming paper.

We find that metal ions decouple from the passive plasma and start to
accelerate steeply at the decoupling radius, whereas the passive
plasma starts to decelerate at this location. The decoupling point from the 
simulation agrees roughly with the
prediction from expression \eqref{decouplerd}, $\rd \approx 1.6$. (To
derive this value, we use the CAK value for $\vinfty$ from
 Table~\ref{Vysledky} and $\beta=0.8$, which is a good estimate for our
model.) The present decoupling contradicts the results of \cite{KK00},
who obtained a shallow, coupled solution.

The quite unexpected result of \cite{KK00} was analysed by
\citet{O2002}, who performed a linear stability analysis of the
time-dependent hydrodynamic equations to derive perturbation growth
rates and propagation speeds in a two-component stellar wind. These 
authors found that the ion decoupling instability persists for
the \cite{KK00} shallow-wind solution, for long wavelength
perturbations with high temporal growth rate $\approx 10^4
\vl/\Rstar$. It therefore seems that the \cite{KK00} solution is not a
physically valid solution.

\citet{O2002} also show that for the solution obtained by
\cite{KK00}, the ion-decoupling instability has a modest spatial
growth rate, mainly due to the fast speed of perturbation propagation,
which only allows a slight amplification during the time needed to
converge to the steady-state solution. We may speculate that this is
related to how \cite{KK00} indeed found convergence of the
iteration scheme applied to solve the steady-state equations to a
shallow, one-component solution.

As a test, we increased the value of the ion charge to a rather
unrealistic value of $\qi=5$, which artificially increases the Coulomb
coupling. For this case we obtained a stable, one-component flow
solution according to CAK.

After decoupling, the ion velocity gradient is so large that the
associated spectral lines from ions should be optically thin. Because
of this dramatic reduction of the Sobolev optical depth in ions, the
absorption of this highly accelerated material becomes weak and  
the signature of this material in the spectrum is also weak. 
 We must mention that the CAK radiative acceleration given by Eq.
\eqref{zazzry} is overestimated. \citet{B1996} showed that inclusion of the shadowing effect
by photospheric lines to radiative acceleration calculation lowers its value significantly
for B stars with a low-density wind compared with the CAK model.
Thus, the high ion speeds from the above figure would not be observed, and the
observed wind speed is instead that of the passive plasma
\citep{S1992}.

\begin{figure*}[t!]
\scalebox{1.0}{\includegraphics{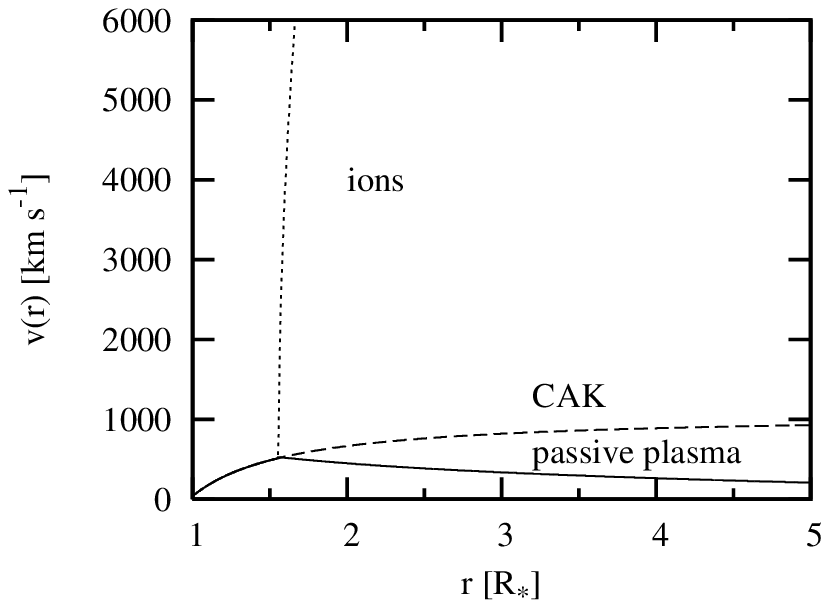}}
\scalebox{1.0}{\includegraphics{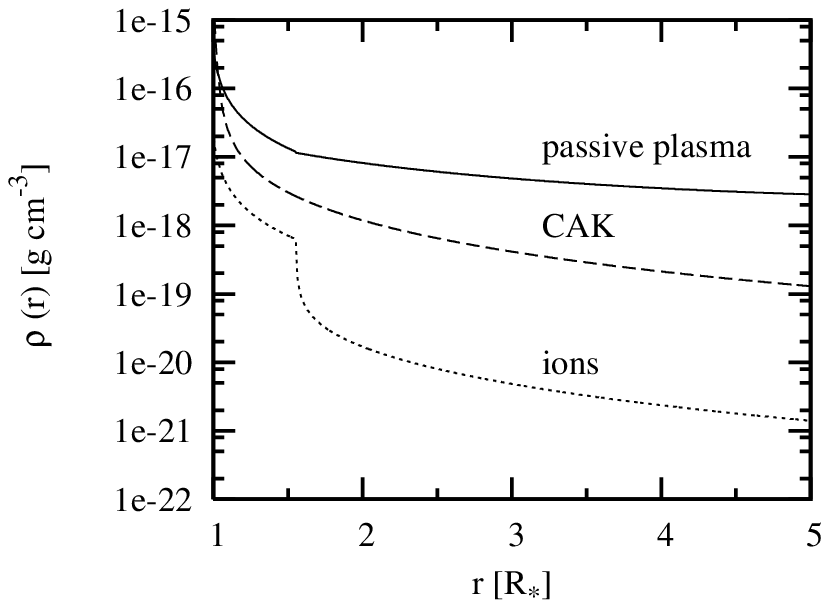}}\\ 
\caption{{\it Left panel:} Final wind-velocity laws for model B5
after 55 flow time units $R_*/\vinfty$. Absorbing ions are marked by
($\cdots$), passive plasma by (---), and the result of the
one-component code by (- - -).  {\it Right panel:} Wind density for
the same model.}
\label{B5}
\end{figure*}

\section{Summary}

We develop a method for time-dependent hydrodynamic simulations of
multi-component stellar winds. To avoid resolving the prohibitively
short friction time scale (causing very stiff system of
equations) in our time-explicit scheme, we used a three-interval fit to
the Chandrasekhar function and solved the friction terms (after
operator-splitting) analytically. The stability and accuracy of our
method is demonstrated for a B0 model star with a well-coupled wind,
where our flow solution evolves to the well-known steady mCAK
solution. 

For a B5 model star with a low-density wind, we find that ion decoupling
from the passive wind plasma occurs at a definite wind radius, instead
of the transition to a shallow, coupled wind solution with small
acceleration of both components, as predicted by \cite{KK00}.

In the future, we will apply the code to a larger sample of stars and
study whether pulsating shells may originate in multi-component stellar
winds \citep{P1999}.

\begin{acknowledgements}
We thank Robert Nikutta and Ji\v{r}\'{\i} Krti\v{c}ka for fruitful
discussions and comments. This work was supported by grant D/04/25764
from the Deutscher Akademischer Austausch Dienst and grant B301630501 from
the Grant Agency of the Academy of Sciences of the Czech Republic, as well
as by Deutsche Forschungsgemeinschaft under grant numbers FE 573/2 and
FE 573/3. The Astronomical Institute Ond\v{r}ejov is supported by 
project AV0\,Z10030501.
\end{acknowledgements}

\end{document}